\documentclass[aps,prl,twocolumn,showpacs,groupedaddress]{revtex4}
\usepackage{graphicx}  
\usepackage{dcolumn}   
\usepackage{bm}        
\usepackage{amssymb}   
\usepackage{verbatim}

\usepackage{amsmath}
\usepackage{amsfonts}

\begin{document}

\title{Calculations for Magnetism in SQUIDs at Millikelvin Temperatures}
\author{S. Sendelbach$^1$}
\author{D. Hover$^{1}$}
\author{A. Kittel$^{2}$}
\author{M. M\"{u}ck$^{3}$}
\author{John M. Martinis$^{4}$}
\author{R. McDermott$^{1,}$}
\email[Electronic address: ]{rfmcdermott@wisc.edu}

\affiliation{$^{1}$Department of Physics, University of
Wisconsin-Madison, Madison, Wisconsin 53706, USA}

\affiliation{$^{2}$Institut f\"{u}r Physik, Carl von Ossietzky
Universit\"{a}t, D-26111 Oldenburg, Germany}

\affiliation{$^{3}$Institut f\"{u}r Angewandte Physik, Justus-Leibig-Universit\"{a}t Gie{\ss}en,
D-35392 Gie{\ss}en, Germany}

\affiliation{$^{4}$Department of Physics, University of California,
Santa Barbara, California 93106, USA}

\date{\today}

\begin{abstract}
Here we present details of a calculation that allows us to extract a surface density of unpaired spins from flux \textit{vs.} temperature experiments performed on field-cooled dc Superconducting QUantum Interference Devices (dc SQUIDs).
\end{abstract}

\maketitle

Our recent measurements show an unexpected dependence of SQUID flux on bath temperature in the millikelvin temperature range \cite{McDermott}. The flux change scales as $1/T$ as temperature is lowered.  This behavior has been observed in both Al and Nb devices, made both with and without a wiring dielectric, and prepared in different facilities according to different fabrication recipes \cite{fab}. Paramagnetic impurities in the materials of the SQUID would naturally give rise to such a signature, and we interpret the $1/T$ dependence of the flux through the SQUIDs as strong evidence for unpaired spins, most likely in the native oxides of the superconductors.

In order to clarify the source of the temperature-dependent flux, we have performed a series of field-cool experiments in which a magnetic field $B_{fc}$ is applied to a 350 pH square-washer Nb SQUID (with inner dimension 200 $\mu$m and outer dimension 1 mm) as it is cooled through $T_c$; the field cool freezes magnetic flux vortices into the Nb film, with density  $\sigma_v\approx B_{fc}/\Phi_0$ \cite{Stan}. When the device is well below $T_c$, the magnetic field is removed, and the SQUID is maintained in a flux-locked loop as it is cooled to millikelvin temperatures. In Fig. 1a we plot the flux threading the SQUID as a function of temperature for eight different values of the cooling field. The cooling field strongly affects the temperature-dependent flux, enhancing or even reversing the polarity of the observed signal. In Fig. 1b we plot the flux change on cooling from 500 mK to 100 mK as a function of the cooling field; a linear fit to the data yields a slope of 1.3 $\Phi_0$/mT. Clearly, vortices contribute significantly to the measured temperature-induced flux shift.

\begin{figure}[b]
\includegraphics[width=.47\textwidth]{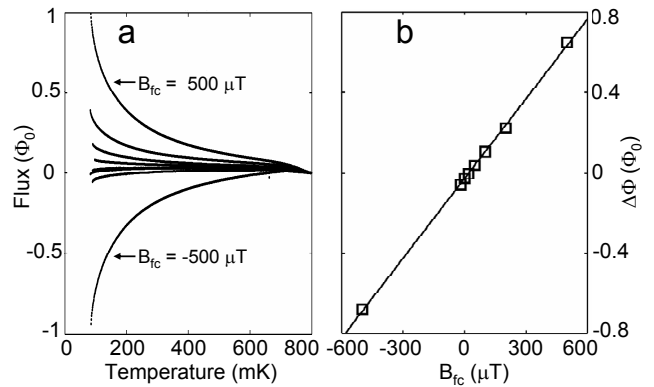}
\vspace*{-0.0in} \caption{(a) Temperature dependence of the flux threading a 350 pH Nb/AlOx/Nb SQUID, for different fields $B_{fc}$ applied as the device was cooled through the superconducting transition. (b) Temperature-induced flux change $\Delta \Phi$ on cooling from 500 mK to 100 mK, as a function of cooling field $B_{fc}$. A linear fit to the data yields a slope $\Delta \Phi/B_{fc}$ = 1.3 $\Phi_0$/mT.}
\label{fig:figure3}\end{figure}
The linear dependence of flux drift on vortex density suggests the following interpretation of the data. As the superconducting films are cooled through $T_c$ in an applied field, vortices nucleate and relatively large magnetic fields are frozen into the films. The observed signal is due to the magnetization of unpaired electron spins on the surface of the superconductor in the strong fields produced by the trapped vortices. For our Nb thin films we expect magnetic fields in the vortex to be of the order of 10 mT. This field strength yields characteristic temperatures of order 10 mK for single electron spins, compatible with the low energy scale seen in the experiments.

Careful analysis of our field-cool data allows us to extract the surface density $\sigma_s$ of spins, a key parameter in models of $1/f$ flux noise from surface magnetism \cite{Koch, de Sousa, Faoro}. We note that there is negligible \textit{direct} coupling to the SQUID from spins polarized out of the plane of the superconducting films; this is easily understood from reciprocity, as surface magnetic fields due to currents in the SQUID have vanishing perpendicular component. However, the topology of our device is quite different from that of a continuous superconducting washer, due to the presence of the vortices. Indeed, the thermal polarization of unpaired surface spins in the vortex forces a redistribution of the circulating supercurrents in the vortex, due to the requirement to conserve magnetic flux. These currents, in turn, couple strongly to the SQUID loop. Calculation of the spin density from the data of Fig. 1 therefore proceeds in two stages: (1) calculation of the coupling between a vortex and the SQUID loop, and (2) calculation of the coupling between surface spins and a vortex. For both parts of this calculation we need to solve for the currents in a superconducting washer with radial symmetry (in the case of the SQUID this is a simplifying assumption); therefore, we first describe the numerical solution to this problem.

This note is organized as follows. In section I, we describe calculation of the current distribution in a thin superconducting washer. In section II, we discuss the coupling of a vortex to a SQUID. In section III, we describe the coupling of surface spins to a vortex. Finally, in section IV, we combine the results of sections II and III to extract a surface density of spins from the data of Fig. 1.

\section{I. Current Distribution in a Superconducting Washer}

We consider a thin circular washer with inner radius $r_i$ and outer radius $r_o$.  The washer is broken into a set of concentric loops with width and turn-to-turn spacing $w$. The vector potential $\vec{A}(x)$ at radius $x$ and produced by a current $I$ through a loop at radius $r$ is given by
\begin{align}
\vec{A}(x) &= \frac{\mu_0}{4\pi} I \oint \frac{r d\vec{\theta}}{
\sqrt{(x-r\cos\theta)^2+(r\sin\theta)^2}}\, .
\end{align}
The differential term $d\vec{\theta}$ has magnitude $\sin\theta$ in the direction parallel to the radius vector of $x$ and $\cos\theta$ in the tangential direction.  Upon integration, only the tangential direction is non-zero.  The flux enclosed in a loop at radius $x$ is given by
\begin{align}
\phi(x) &=2\pi x |\vec{A}(x)| \\
&=\pi x \mu_0  I A_c(x/r) \ , \\
A_c(x/r) &= \oint \frac{ \cos\theta\ d\theta/2\pi}{
\sqrt{(x/r-\cos\theta)^2+\sin^2\theta}}
\nonumber\\
&=\frac{1}{\pi \sqrt{x/r}}\left[\left(\frac{2}{k}-k\right)K(k)-\frac{2}{k}E(k)\right],
\end{align}
where we have defined
\begin{align}
k\equiv \sqrt{\frac{4x/r}{\left(1+x/r\right)^2}}\, ,
\end{align}
and where $K(k)$ and $E(k)$ are the complete elliptic integrals of the first and second kind, respectively. The functional dependence of $A_c$ is plotted in Fig. 1.

\begin{figure}[t]
\includegraphics[width=.47\textwidth]{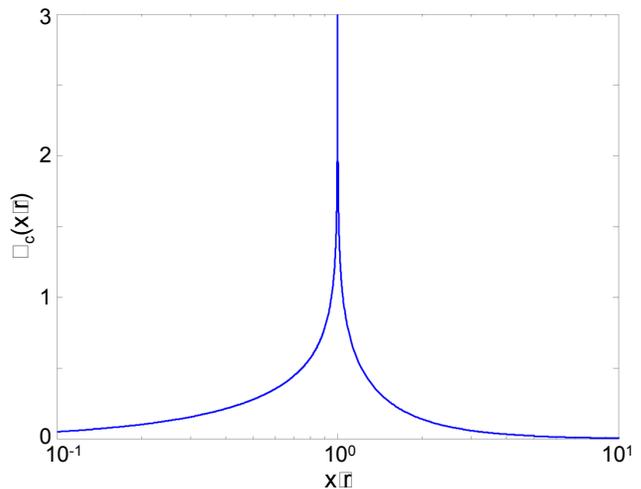}
\vspace*{-0.0in} \caption{Plot of normalized magnetic potential $A_c$ versus normalized distance $x/r$.}
\label{fig:Ac}\end{figure}

The self and mutual inductances between loops $i$ at radius $x$ and $j$ at radius $r$ are given by
\begin{align}
M_{ij}=\pi\mu_0 x \frac{A_c([x+w/2]/r)+A_c([x-w/2]/r)}{2} \ ,
\end{align}
where we have approximately averaged over the width of the loop by evaluating $A_c$ at $x\pm w/2$.  The currents in loops 1 to $N$ are described by a vector $\vec{i}$; the induced flux $\vec{\phi}$ in the loops is then given by
\begin{align}
\vec{\phi} = M \vec{i} \ .
\end{align}
In the case of a superconducting washer with vanishingly small London penetration depth $\lambda_L$, the flux quantization condition
\begin{align}
n\Phi_0 = \displaystyle \oint \,\textbf{A}\cdot d\textbf{r}
\end{align}
requires a constant distribution of fluxes through the washer. The current distribution is determined by setting $\vec{\phi} = \Phi_0$ for all loops and multiplying by the inverse of the mutual inductance matrix:
\begin{align}
\vec{i} = M^{-1} \vec{\phi} \ .
\label{eq:invert}
\end{align}

\section{II. Coupling of a Vortex to the SQUID}

A single vortex trapped in the SQUID washer induces an effective flux offset in the SQUID.  The magnitude of this effect can be understood by considering a few simple examples.  First, for a vortex trapped in an infinite superconducting sheet, the phase change between points above and below the vortex is $\pi$, corresponding to a $\Phi_0/2$ flux shift.  By symmetry, this phase change is unchanged if the vortex is centered in a superconducting strip of finite width.  A vortex in a SQUID with a narrow washer $(r_i-r_o) \ll r_o$ will similarly produce a flux offset of $\Phi_0/2$ in the SQUID if the vortex is placed in the center of the wire at $(r_i+r_o)/2$.  For a vortex placed near the inner radius $r_i$, the flux offset will approach $1\ \Phi_0$, whereas a vortex placed near the outer radius $r_o$ will give a flux offset approaching $0\ \Phi_0$.  For vortices trapped uniformly throughout the narrow washer, the average flux offset will be $\Phi_0/2$.

To calculate the coupling between a SQUID and a vortex trapped in the SQUID washer at radius $r$, we consider a radially symmetric geometry that is amenable to the calculation technique described above. From symmetry, the flux coupling is independent of angular orientation of the vortex around the washer.  Thus, the flux coupling from one vortex is equivalent to $n$ vortices of magnitude $\Phi_0/n$ placed around the washer with uniform angular spacing and identical radius.  As $n \rightarrow \infty$, the vortices are equivalent to a cut in the washer with a net flux through the cut of magnitude $\Phi_0$.  This problem can now be solved using the matrix inversion technique of eq. \ref{eq:invert}.  For the solution, we need to calculate the effective flux $f$ that must be applied to the SQUID so that the total circulating current is unchanged when a flux quantum is applied to the cut.  The flux profile in the loops is thus
\begin{align}
\phi(r') &= -f,  &r'<r;
\nonumber\\
\phi(r') &= \Phi_0-f,  &r'>r.
\end{align}
The offset flux $f$ is solved using eq. \ref{eq:invert} with the constraint that the total circulating current $\displaystyle \sum i$ is zero.  The solution of this problem $f(r)$ is shown in Fig. 3 for both a narrow and wide washer.

By solving for $f(r)$ over the range $r_i\leq r \leq r_o$ and integrating over the area of the SQUID, we compute the average flux $\langle f(r_i,r_o) \rangle $ coupled to the SQUID from a vortex with a uniform distribution across the washer:
\begin{align}
\langle f(r_i,r_o) \rangle \equiv
\frac{\int^{r_o}_{r_i} r f(r) \,dr}{(r_o^2-r_i^2)/2 } .
\end{align}
Numerical calculations for a circular washer show that $\langle f \rangle = 0.5\,\Phi_0$ for a narrow washer, as expected; for our device dimensions (inner radius 100 $\mu$m and outer radius 500 $\mu$m), we find a coupling factor $\langle f \rangle = 0.14\,\Phi_0$.

\begin{figure}[t]
\includegraphics[width=.47\textwidth]{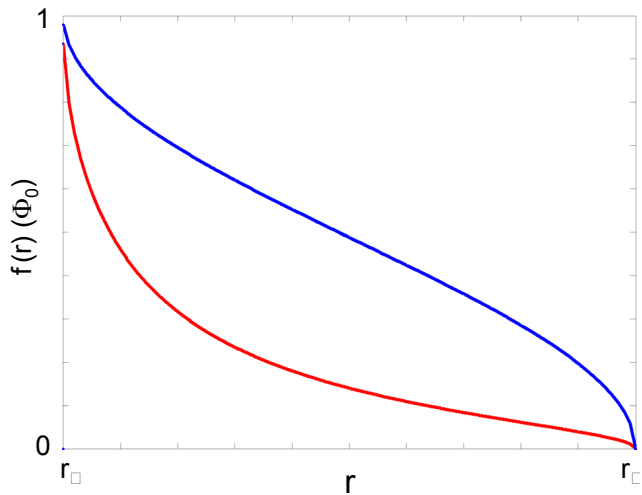}
\vspace*{-0.0in} \caption{Coupling $f(r)$ of a vortex to a SQUID: (Blue) $r_o/r_i$ = 1.02; (Red) $r_o/r_i$ = 5.  The average flux offset for the thin washer is $\langle f\rangle = 0.5\,\Phi_0$, whereas for the large washer we find $\langle f\rangle = 0.14\, \Phi_0$.}
\label{fig:I_Phi}\end{figure}

What is the physical source of the offset flux?  Vortex currents, although localized mostly around the core, fall off over a long distance.  Part of the current must then circulate around the center hole of the SQUID washer, giving an effective flux offset to the SQUID loop.

\section{III. Coupling of Surface Spins to a Vortex}

The total magnetic flux through a vortex is quantized, and has contributions from magnetic fields generated by the currents circulating around the vortex along with any externally applied magnetic field.  This external magnetic field is typically considered to be zero because of screening from the superconductor.  However, the surface spins located within the vortex are strongly coupled to the vortex, so their magnetic flux cannot be neglected. As these spins become polarized, they couple a flux $\Phi_v(T)$ to the vortex; the quantization condition requires a redistribution of currents in the vortex. The change in average flux coupled to the SQUID is thus $\Phi = -0.14\,\Phi_v$ per vortex for the thick washer geometry consider previously.

To calculate $\Phi_v(T) $ we rely on reciprocity: if we know the current distribution in the vortex, we can calculate the magnetic fields at all points in space. This allows us to calculate the coupling to a spin at an arbitrary location; the total flux is obtained by integrating over the area of the vortex. We first discuss calculation of the current distribution in the vortex, then we describe the coupling of the vortex to surface spins.

We model the vortex as a superconducting washer with inner radius $r_i = \xi$ and outer radius $r_o>>\xi$, where $\xi$ is the coherence length. The current distribution is calculated using the techniques described in section I. In this case, however, we have a penetration depth $\lambda_L$ that is larger than the inner dimensions of the washer, and flux is not constant across the superconductor. Instead, single-valuedness of the superconducting wavefunction imposes the following quantization condition:
\begin{align}
\Phi_0 = \displaystyle \oint \,\left(\textbf{A}+\mu_0 \lambda_L^2 \textbf{J}\right)\cdot d\textbf{r},
\end{align}
where the integral on the right-hand side is London's fluxoid. The new term adds additional elements on the diagonal of the inductance matrix:
\begin{align}
\left(M_{Lon}\right)_{ij} &= \pi \mu_0 x \frac{A_c\left(\left[x+w/2\right]/r\right)+A_c\left(\left[x-w/2\right]/r\right)}{2}
\nonumber\\
        &+ \delta_{ij}\, 2\pi \mu_0 x \left(\frac{\lambda_L^2}{dw}\right),
\label{eq:Mlon}
\end{align}
where $\delta_{ij}$ is the Kronecker $\delta$-function and $d$ is the film thickness. Inversion of the matrix $M_{Lon}$ allows determination of the current profile in the vortex. A two-dimensional solution of the vortex shows that, for the materials parameters we are interested in here, the currents are uniform along the $z$ direction to within about 10$\%$. Therefore we may simplify the matrix inversion part of the calculation by using 1-D discretization and replacing $2\lambda_L^2/dw$ in eq. \ref{eq:Mlon} by $\Lambda/w$, where $\Lambda \equiv 2\lambda_L^2/d$ is the thin-film penetration depth.  In Fig. 4 we plot the current and flux distributions in the vortex for $\xi$ = 30 nm and $\Lambda$ = 100 nm.

\begin{figure}[t]
\includegraphics[width=.47\textwidth]{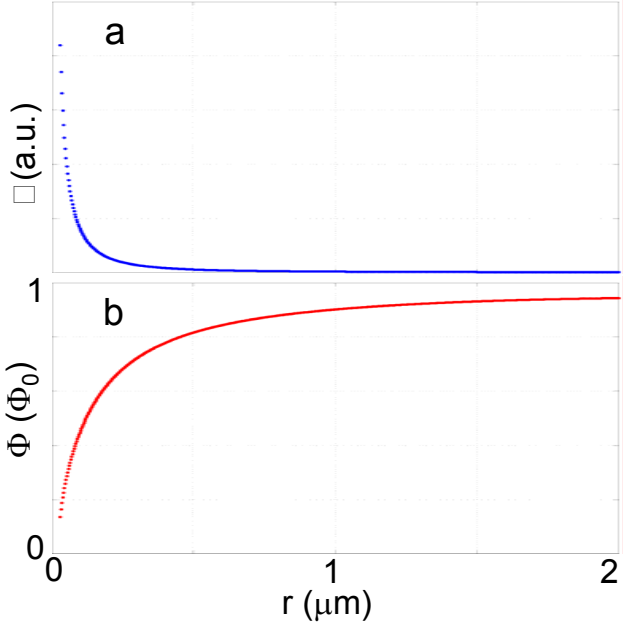}
\vspace*{-0.0in} \caption{(a) Current density and (b) flux distribution in a vortex with $\xi$ = 30 nm and $\Lambda$ = 100 nm.}
\label{fig:I_Phi}\end{figure}
The vortex self-inductance $L_v$ is easily determined from the total current $I_v = \displaystyle \sum i$ in the washer:
\begin{align}
L_v = \Phi_0/I_v.
\end{align}
For $\xi$ = 30 nm and $\Lambda$ = 100 nm, we find a vortex self-inductance $L_v$ = 0.24 pH.

Once the currents in the vortex are known, it is straightforward to calculate the in-plane $B_r(r,z)$ and out-of-plane $B_z(r,z)$ magnetic fields at all points in space by summing over the contributions $b_{r,z}(r,z)$ of the individual loops of current $i$. For a current loop with radius $a$ centered at $r=0$ in the $z=0$ plane, we have
\begin{align}
b_r(r,z) &= -\frac{\mu_0ikz}{4\pi\sqrt{ar^3}}\left(K(k)-\frac{2-k^2}{2\left(1-k^2\right)}E(k)\right);
\nonumber\\
b_z(r,z) &= \frac{\mu_0ikr}{4\pi\sqrt{ar^3}}\left(K(k)+\frac{k^2\left(r+a\right)-2r}{2r\left(1-k^2\right)}E(k)\right),
\end{align}
where now we have
\begin{align}
k \equiv \sqrt{\frac{4r/a}{(1+r/a)^2+(z/a)^2}},
\end{align}
and where again $K(k)$ and $E(k)$ are the complete elliptic integrals of the first and second kind, respectively.

\begin{figure}[b]
\includegraphics[width=.47\textwidth]{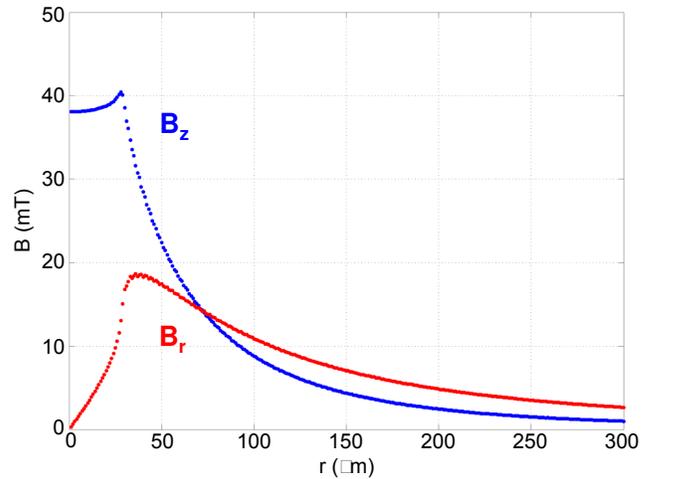}
\vspace*{-0.0in} \caption{Distribution of in-plane $B_r$ and out-of-plane $B_z$ magnetic fields for a vortex with $\xi$ = 30 nm and $\Lambda$ = 100 nm. Here we consider a film with thickness $d$ = 80 nm and evaluate the fields on the surface of the film, assuming a uniform current density throughout the thickness of the film.}
\label{fig:BrBz}
\end{figure}
We now consider magnetic coupling between a vortex and a surface spin located at radius $r$. We first think of the spin as an infinitesimal current loop with effective sensing area $A_{eff}$. The flux coupled to the spin by the vortex is thus
\begin{equation}
\Phi_{v\rightarrow s} = B(r)A_{eff},
\label{eq:coupling1}
\end{equation}
where $B(r)=[B_z(r)^2+B_r(r)^2]^{1/2}$ is the magnitude of the field at $r$. Note that the geometrical coupling of the spin to the vortex is unity, since the spin is polarized in the direction of the local field. The mutual inductance of the spin to the vortex is then
\begin{align}
M &= \frac{B(r)A_{eff}}{I_v}
\nonumber\\
\nonumber\\
  &=L_v\frac{B(r)A_{eff}}{\Phi_0}
\label{eq:mut}
\end{align}

The flux coupled to the vortex by a spin at $r$ with moment $m=\mu_B \tanh \left(\mu_B B(r)/2k_BT\right)$ is thus
\begin{align}
\Phi_{s\rightarrow v} = \mu_B L_v \frac{1}{\Phi_0} B(r) \tanh \left(\frac{\mu_B B(r)}{2k_BT}\right).
\label{eq:coupling2}
\end{align}
The total flux coupled to the vortex by spins with surface density $\sigma_s$ is therefore
\begin{align}
\Phi_v(T) &= \mu_B \sigma_s L_v \frac{1}{\Phi_0} \displaystyle \int \,2\pi r B(r) \tanh \left(\frac{\mu_B B(r)}{2k_BT}\right)\,dr
\nonumber\\
\nonumber\\
        &= \mu_B \sigma_s L_v P_{eff}(T),
\label{eq:phitot}
\end{align}
where we have introduced the effective spin polarization $P_{eff}(T)$, defined as
\begin{align}
P_{eff}(T) \equiv \frac{1}{\Phi_0} \displaystyle \int \,2\pi r B(r) \tanh \left(\frac{\mu_B B(r)}{2k_BT}\right)\,dr.
\label{eq:Peff}
\end{align}
From the known current distribution in the vortex we can calculate $P_{eff}(T)$. Note that in eq. \ref{eq:Peff} we are also integrating the current over the thickness of the washer.  Because we have a uniform current distribution in $z$, we average the fields over the thickness of the washer.


\section{IV. Analysis of Field-Cool Data}

As the SQUID is cooled to millikelvin temperatures, a change in temperature yields a change in effective spin polarization in the vortex, which in turn induces a change in flux coupled to each vortex. The total flux change $\Phi(T)$ at the SQUID is obtained by summing over all vortices, taking into account the coupling factor 0.14. We find
\begin{align}
\Phi(T) = 0.14\,A_{SQ}\,\sigma_v \, \mu_B \sigma_s L_v \,P_{eff}(T),
\label{eq:totalFlux}
\end{align}
%
where $A_{SQ}$ is the area of the SQUID washer, and where $\sigma_v$ is the density of vortices in the SQUID washer. For our field-cool experiments, we have $\sigma_v \approx B_{fc}/\Phi_0$, where $B_{fc}$ is the magnitude of the cooling field. In these experiments, we see a linear dependence of flux drift $\Delta \Phi \equiv \Phi \left(100 \, \textrm{mK}\right) - \Phi \left(500 \, \textrm{mK}\right)$ on vortex density. A linear fit to the data of Fig. 1b yields a slope $\Delta \Phi/B_{fc} = 1.3 \Phi_0$/mT. This slope is related to spin density as follows:
\begin{align}
\frac{\Delta \Phi}{B_{fc}} = 0.14 \frac{A_{SQ}}{\Phi_0} \, \mu_B \sigma_s L_v \,\Delta P_{eff},
\label{eq:totalFlux}
\end{align}
where $\Delta P_{eff} \equiv P_{eff}\left(100 \, \textrm{mK}\right) - P_{eff}\left(500 \, \textrm{mK}\right)$. For our materials parameters, we find $\Delta P_{eff}$ = 0.037. Using $A_{SQ}$ = 0.96 mm${^2}$ and the slope $\Delta \Phi/B_{fc}$ = 1.3 $\Phi_0$/mT from Fig. 1b, we extract a spin density $\sigma_s$ = $5.0 \times 10^{17}$ m$^{-2}$. This density of surface spins is compatible with densities considered in recent theoretical models of $1/f$ flux noise from surface spins \cite{Koch, de Sousa, Faoro}.

\begin{acknowledgments}
We acknowledge useful discussions with L. Faoro, L.B. Ioffe, B.L.T. Plourde, and C.C. Yu. Some devices were fabricated at the UCSB Nanofabrication Facility, part of the NSF-funded NNIN. This work was supported in part by the U.S. Government. The views and conclusions contained in this document are those of the authors and should not be interpreted as representing the official policies, either expressly or implied, of the U.S. Government.
\end{acknowledgments}


\begin{thebibliography}{99}

  \bibitem{McDermott}
S. Sendelbach \textit{et al.}, arXiv:0802.1518.

  \bibitem{fab}
The Al-AlOx-Al SQUIDs were fabricated at UC Santa Barbara on oxidized silicon substrates, and comprised sputter-deposited Al films, thermally-grown AlOx, PECVD-grown SiO$_2$, and evaporated CuAu shunt resistors. The Nb-AlOx-Nb SQUIDs were fabricated at the University of Gie{\ss}en on oxidized silicon or sapphire substrates, and comprised sputtered Nb, sputtered Al, thermally grown AlOx, and sputtered Pd shunt resistors; in addition, some of the Nb-AlOx-Nb devices included a sputtered SiO wiring dielectric.

  \bibitem{Stan}
G. Stan \textit{et al}.,  Phys. Rev. Lett. \textbf{92}, 097003 (2004).

 \bibitem{Koch}
R.H. Koch \textit{et al}., Phys. Rev. Lett. \textbf{98}, 267003 (2007).

  \bibitem{de Sousa}
R. de Sousa, Phys. Rev. B \textbf{76}, 245306 (2007).

  \bibitem{Faoro}
L. Faoro and L.B. Ioffe, arXiv:0712.2834 (2007).

\end{thebibliography}
\end{document}